\documentclass[a4paper]{article}

\usepackage[english]{babel}
\usepackage[utf8x]{inputenc}
\usepackage[T1]{fontenc}
\usepackage{booktabs}
\usepackage{amsfonts}

\usepackage[a4paper,top=3cm,bottom=2cm,left=3cm,right=3cm,marginparwidth=1.75cm]{geometry}

\usepackage{amsmath}
\usepackage{graphicx}
\usepackage{mathtools} 
\usepackage[colorinlistoftodos]{todonotes}
\usepackage[colorlinks=true, allcolors=blue]{hyperref}
\usepackage{subcaption}

\usepackage{bm}
\usepackage{algorithm} 
\usepackage{algpseudocode}
\usepackage{amssymb,amsmath,amsthm}
\usepackage{authblk}

\usepackage[shortcuts]{extdash} 

\newcommand{\atan}{\qopname\relax o{atan2}}

\title{Secure protocol to protect location privacy in distance calculation}
\author{Cristina Romero-Tris}
\author{David Meg\'ias}
\affil{Internet Interdisciplinary Institute (IN3) \\
  Universitat Oberta de Catalunya (UOC) \\
  CYBERCAT-Center for Cybersecurity Research of Catalonia\\
  Email: \{cromerotr,dmegias\}@uoc.edu}

\begin{document}
\maketitle

\begin{abstract}

Several applications require computing distances between different people. For example, this is required if we want to obtain the close contacts of people in case of and epidemic, or when restraining orders are imposed by a judge.
However, periodically revealing location might pose a privacy threat to the involved parties. Continuous location data may be used to infer personal information about the owner, like behaviors, religious beliefs, buying habits, routines, etc. In this paper, we show that it is possible to calculate distance between two parties without disclosing their latitude and longitude data. For this purpose, we design a secure protocol based on the 
ElGamal cryptosystem and its homomorphic properties. The proposed protocol allows the 
calculation of distances while preserving location privacy. The protocol is analyzed in terms of security and performance. The security analysis shows that no involved party can learn any information about location. 
\end{abstract}

\section{Introduction}

The continuous growth of mobile applications has lead to a high number of services which require information about users' location. In this context, we find several applications which require computing distances between two individuals. One example of these applications is in case of an epidemic, where a central entity may need to control distances within the population. Another example could be in the case of restraining orders imposed by a judge, where a central entity needs to verify the distance between a victim and its aggressor. 

Even if these services make a real contribution to the community, they can also pose a serious threaten to users' privacy. The straightforward way to find the distance between two users requires that both reveal their actual location. Based on their right of privacy, users might be unwilling to reveal this information. For this reason, it is necessary to provide alternatives which allow users to benefit from these location-based services without giving up their privacy.

Encryption is a strong asset to provide data privacy. However, while encryption can secure data in transit and stored, to put data to use at some point it needs to be decrypted, and in that moment it becomes vulnerable. However, secure multi-party computation \cite{damgaard10} is an area of cryptography which provides the ability to compute values of interest from multiple encrypted data sources without any party having to reveal their private data. Secure multi-party protocols are applied in different areas \cite{yao82}, like secret voting, or database querying. In these scenarios, users compute the value of a function without unduly giving away any information about the values of their own variables.

Similarly to secure multi-party computation, we present a cryptographic protocol based on multiplicative homomorphic encryption. The proposed protocol makes it possible to calculate distance between 
two entities without revealing their actual coordinates. 

Regarding the distance calculation, we employ the Haversine formula \cite{sinn84}. This formula determines the great-circle distance between two points on a sphere given their longitudes and latitudes. Regarding cryptography, we use the ElGamal \cite{ElGamal85} cryptosystem and its homomorphic properties.


\section{Contributions and plan of this paper}
In this paper, we address the situation where two users need to learn the distance between them, but do not want to reveal their exact location. For this purpose, we present a cryptographic protocol based on ElGamal cryptosystem and its homomorphic properties to calculate the Haversine distance between two points without revealing their GPS coordinates.


The rest of the paper is organized as follows. Section \ref{sec:background} describes the background tools used in our protocol. Section \ref{sec:related} summarizes some literature related to our work. Section \ref{protocol} describes the proposed protocol in detail. Sections \ref{sec:security} and \ref{sec:performance} analyze the protocol in terms of security and performance. Finally, Section \ref{sec:conclusions} presents the conclusions of the work.

\section{Background} \label{sec:background}
As mentioned before, the designed cryptographic protocol is based on homomorphic properties. In order 
to calculate distance, we employ the Haversine formula. In this section, we describe these components which are later used in the protocol. 

\subsection{ElGamal cryptosystem}
Details and proofs of ElGamal cryptosystem can be found in \cite{ElGamal85}. However, for the sake of notation in later sections, we summarize the main operations:

\begin{itemize}
\item{\emph{Key generation}} \label{key_generation}

First, a large random prime number $p$ is generated, where $p=2q+1$ and $q$ is a 
prime number too. Also, a generator $g$ of the multiplicative group $\mathbb{Z}^*_q$ 
is chosen. Then, the user generates a random private secret key $s \in \mathbb{Z}^*_q$ and calculates the public key $y=g^{s}$.

\item{\emph{Message encryption}}

Given a message $m$ and a public key $y$,
a random value $r$ is generated and the ciphertext is computed as follows:

\begin{equation}
E_y(m)=c=(c1,c2)=(g^r, m \cdot y^r).
\end{equation} 

\item{\emph{Message decryption}}\label{sec:partial}

A cyphertext  $E_{y}(m)=(c1,c2)$ is decrypted using the private key as follows:
  
\begin{equation}
  m=\frac{c2}{c1^{s}}=\frac{m \cdot y^r}{{g}^{r\cdot s}}
\end{equation}   
  
\end{itemize}

\subsubsection{Multiplicative homomorphic ElGamal}
Homomorphic encryption schemes are used to perform operations on ciphertexts without decrypting data. When the result of the operation is decrypted, it is the same as if the calculation had been carried out on raw data. 

More precisely, a (group) homomorphic encryption scheme over a group $(G, ∗)$ satisfies that given two encryptions $c_1 = E_k(m_1)$ and $c_2 = E_k(m_2)$, where $m_1$, $m_2 \in G$ and $k$ is the encryption
key, one can efficiently compute $E_k(m_1 ∗ m_2)$ without decrypting $c_1$ and $c_2$.

ElGamal cryptosystem is a multiplicative homomorphic encryption scheme. This means that the generic operation "$∗$" described above is the product operation "$\cdot$". 

A proof of the ElGamal cryptosystem homomorphic properties can be found in \cite{liu16}. Nevertheless, it can be easily seen that, for a public key $y$, two random numbers $r_1$ and $r_2$, and two messages $m_1$ and $m_2$:
\begin{equation}
E_y(m_1) \cdot E_y(m_2) = (g^{r_1}, m_1 \cdot y^{r_1}) \cdot (g^{r_2}, m_2 \cdot y^{r_2}) = (g^{r_1 + r_2}, (m_1 \cdot m_2) \cdot y^{r_1+r_2}) = E_y(m_1 \cdot m_2) 
\end{equation} 

This simply means that performing a product between two ElGamal ciphertexts, we obtain the encrypted version of their cleartext product. This is a useful property that we will employ in the proposed protocol.

\subsection{The Haversine formula} \label{sec:haversine}

The Haversine formula started coming into use in the beginning of the 19th century, for navigation purposes \cite{sinn84}. This formula gives minimum distance between any two points on a spherical body by using latitude and longitude.  It is a special case of a more general formula in spherical trigonometry, the law of haversines, that relates the sides and angles of spherical triangles.

Assuming that we want to calculate distance between user 1 located at latitude and longitude $(\lambda_1, \varphi_1)$, and user 2 located at $(\lambda_2, \varphi_2)$, the Haversine distance $d$ is:

\begin{equation}
d=2 R  \arcsin \left ( \sqrt{\sin^2\left ({\frac{\lambda_2-\lambda_1}{2}}\right ) + \cos \lambda_1 \cos \lambda_2  \sin^2\left ({\frac{\varphi_2-\varphi_1}{2}}\right )} \right )
\end{equation} 

where R is the Earth radius, i.e.,  the approximate distance from Earth's center to its surface, about 6,371 km. 

For simplification, the Haversine formula is often expressed as a list of steps:
\begin{equation}
\begin{split}
&\Delta\lambda =  \lambda_2 - \lambda_1 \\
&\Delta\varphi =  \varphi_2 - \varphi_1 \\
&a=\sin ^2\left(\frac{\Delta\lambda}{2}\right)+\cos \lambda_1 \cos \lambda_2 \sin^2\left(\frac{\Delta\varphi}{2}\right)\\
&d=2R \atan\left(\sqrt{a}, \sqrt{1-a}\right)
\end{split}
\end{equation}
where $\atan(\cdot,\cdot)$ is the 2-argument variant of the arctangent function.  

Nevertheless, since we will be using a multiplicative homomorphic ElGamal in our protocol, it is more convenient to express all the calculations related to $\lambda_1, \varphi_1, \lambda_2, \varphi_2$ as products. In order to do so, we apply the trigonometric conversion for the angle difference identity:

\begin{equation}
\sin (\alpha-\beta) = \sin \alpha  \cos \beta - \cos \alpha  \sin \beta 
\end{equation}

Consequently, we can express the Haversine formula as the following list of steps:
\begin{equation} \label{eq:haver}
\begin{split}
&i=\cos\left(\frac{\lambda_1}{2}\right)  \sin\left(\frac{\lambda_2}{2}\right) -  \sin\left(\frac{\lambda_1}{2}\right) \cos\left(\frac{\lambda_2}{2}\right)\\
&j= \cos\left(\frac{\varphi_1}{2}\right)  \sin\left(\frac{\varphi_2}{2}\right) -  \sin\left(\frac{\varphi_1}{2}\right) \cos\left(\frac{\varphi_2}{2}\right) \\   
&a=i^2+j^2 \cos \lambda_1 \cos \lambda_2  \\
&d=2R \atan\left(\sqrt{a}, \sqrt{1-a}\right)
\end{split}
\end{equation}

Now, by rearranging terms, we can write the following formulae:

\begin{equation} \label{eq:expr_a}
\begin{split}
i_1=&\cos\left(\frac{\lambda_1}{2}\right)  \sin\left(\frac{\lambda_2}{2}\right)\\
i_2=&\sin\left(\frac{\lambda_1}{2}\right) \cos\left(\frac{\lambda_2}{2}\right)\\
j_1=&\cos\left(\frac{\varphi_1}{2}\right)  \sin\left(\frac{\varphi_2}{2}\right)\\
j_2=&\sin\left(\frac{\varphi_1}{2}\right) \cos\left(\frac{\varphi_2}{2}\right)\\
m=&\cos \lambda_1 \cos \lambda_2\\
a=& i_1^2 - 2i_1i_2 +i_2^2+mj_1^2-2mj_1j_2+mj_2^2
\end{split}
\end{equation}

Hence, $a$ can be computed by adding up six terms, each of which can be computed only with multiplications: $a = t_1+t_2+t_3+t_4+t_5+t_6$ with 
\begin{equation}\label{eq:terms}
\begin{split}
 t_1=&i_1^2\\
 t_2=&-2i_1i_2\\
 t_3=&i_2^2\\
 t_4=&mj_1^2\\
 t_5=&-2mj_1j_2\\
 t_6=&mj_2^2  
\end{split}
\end{equation}

Note that, once $a$ is available, the distance $d$ can be directly obtained using the expression for $d$ detailed in Equation \ref{eq:haver}. The values of the different terms are detailed in Table \ref{tab:terms_a}.

\begin{table}[htbp]
  \centering
  \def\arraystretch{2}
  \caption{Terms for the computation of $a$}
  \label{tab:terms_a}
  \begin{tabular}{|l|l|l|}
   \hline
   $t_1=\cos^2\left(\frac{\lambda_1}{2}\right) \sin^2\left(\frac{\lambda_2}{2}\right)$ &
   $t_2=-\frac{1}{2}\sin \lambda_1\sin \lambda_2$ &
   $t_3=\sin^2\left(\frac{\lambda_1}{2}\right) \cos^2\left(\frac{\lambda_2}{2}\right)$ \\ \hline
   $t_4=m\cos^2\left(\frac{\varphi_1}{2}\right) \sin^2\left(\frac{\varphi}{2}\right)$ &
   $t_5=-\frac{1}{2}m\sin \varphi_1\sin \varphi_2$ &
   $t_6=m\sin^2\left(\frac{\varphi_1}{2}\right) \cos^2\left(\frac{\varphi}{2}\right)$ \\
   \hline 
\end{tabular}
\end{table}

An equivalent expression can be found for the terms of $a$ making use of the following trigonometric identity:
\[
\cos^2\alpha\sin^2\beta+\sin^2\alpha\cos^2\beta = \frac{1}{2}-\frac{1}{2}\cos(2\alpha)\cos(2\beta)\]
Hence:
\[
t_1+t_3=\frac{1}{2}-\frac{1}{2}\cos\left(2\frac{\lambda_1}{2}\right)\cos\left(2\frac{\lambda_2}{2}\right)=\frac{1}{2}-\frac{1}{2}m
\]
and
\[
t_4+t_6=\frac{1}{2}m-\frac{1}{2}m\cos\left(\frac{2\varphi_1}{2}\right)\cos\left(2\frac{\varphi_2}{2}\right)=\frac{1}{2}m-\frac{1}{2}m\cos\varphi_1\cos\varphi_2
\]
Thus, $t_1$ and $t_3$ in Table \ref{tab:terms_a} could be replaced by the following terms:
\begin{equation}
\begin{split}
    \tilde t_1&=\frac{1}{2} \\
    \tilde t_3&=-\frac{1}{2}m
\end{split}
\end{equation}
Similarly, both $t_4$ and $t_6$ can be replaced by the following terms:
\begin{equation}
\begin{split}
    \tilde t_4&=\frac{1}{2}m \\
    \tilde t_6&=-\frac{1}{2}m\cos\varphi_1\cos\varphi_2
\end{split}
\end{equation}

Finally, there are at least four equivalent ways to divide $a$ in different terms that, added up together, lead to the same value:
    \begin{equation} \label{eq:1} a=t_1+t_2+t_3+t_4+t_5+t_6\end{equation} 
     \begin{equation}\label{eq:2} a=\tilde t_1+t_2+\tilde t_3+t_4+t_5+t_6 \end{equation} 
     \begin{equation}\label{eq:3} a=t_1+t_2+t_3+\tilde t_4+t_5+\tilde t_6 \end{equation} 
     \begin{equation}\label{eq:4} a=\tilde t_1+t_2+t_5+\tilde t_6
    \end{equation} 
Note that $\tilde t_3$ and $\tilde t_4$ cancel each other. Hence, the computation can be reduced to:
\begin{equation}
    a=\tilde t_1+t_2+t_5+\tilde t_6=\frac{1}{2}+t_2+t_5+\tilde t_6
\end{equation}
with $t_2=-\frac{1}{2}\sin \lambda_1\sin \lambda_2$, $m=\cos \lambda_1 \cos \lambda_2$, $t_5=-\frac{1}{2}m\sin \varphi_1\sin \varphi_2$ and $\tilde a_6=-\frac{1}{2}m\cos\varphi_1\cos\varphi_2$.

To simplify the notation, we refer to three terms $a_1$, $a_2$ and $a_3$ that are required to compute $a$ with $a_1=t_2$, $a_2=t_5$ and $a_3=\tilde t_6$. In principle, if $P_1$ and $P_2$ are not very far and their location angles are relatively similar (at least belonging to the same quadrant), all terms $a_i$ will be negative.

\section{Related work} \label{sec:related}
Secure Multi-Party Computation, often denoted as MPC, allows a set of users $U_1, \ldots, U_n$, who each possess some initial values $x_1, \ldots, x_n$, to securely compute some function $f(x_1, \ldots, x_n) = (y_1, \ldots, y_n)$ such that $U_i$
learns nothing but $y_i$. 

MPC has been applied in the literature in several fields. For example, voting protocols like \cite{balas16, sharma16} need to count votes but maintaining secrecy voting for each user. In a Yes-No voting (Y-N), a simplistic solution is to use an additive homomorphic cryptosystem, like Paillier \cite{paillier}. Each user may encrypt a positive vote as a zero, and a negative vote as a one. Adding all the encrypted votes and then collaboratively decrypting the outcome, will result into the addition of all the positive votes. 

Another field where MPC is applied is the auctions domain \cite{cachin99, boge06}. In this scenario, a set of bidders propose a sealed (encrypted) bid. Using secure multi-party computation, and again homomorphic encryption, it is possible to determine which is the highest bid, without revealing the amounts proposed by the other bidders. 

Benchmarking is another area where MPC can be applied \cite{atallah04, damgaard16}. In this context, there are several companies willing to analyze their business, comparing themselves with other companies in the market. This process has to be done while preserving confidentiality of companies’ private data.

Many of these solutions have been successfully deployed in specific countries. For example, the work presented in \cite{boge09} the authors use MPC for the calculation of prices for the Danish sugar beet market. In \cite{besta17}, MPC has been used to  evaluate gender pay disparities in Boston. Another example is found in \cite{Bogdanov15}, which builds a tax fraud detection system prototype for Estonia, based again on MPC.

In the field of location privacy, MPC has been applied in the work presented in \cite{bilo14}. The proposed solution allows a group of users to find their optimal meeting location. The protocol employs the additive homomorphic properties of Paillier cryptosystem \cite{paillier} and the multiplicative homomorphic properties of ElGamal \cite{ElGamal85}. However, combining two cryptosystems introduces an overload in terms of complexity, computation and communication. Their proposal requires group key generation and storage for each cryptosystem, more encryption and decryption operations and extra message exchanges between users. Besides, their solution computes euclidean distance based on cartesian coordinates, less precise in our scenario since the Earth curvature is neglected. 
Our proposal outperforms this work by calculating the Haversine distance with GPS coordinates, easily obtainable from any mobile device. Besides, our proposal computes distances using only ElGamal cryptosystem, reducing hence the number of keys, complexity and communication.

\section{Proposed protocol} \label{protocol}
\begin{figure}[!ht]
\centering
\includegraphics[width=1\linewidth]{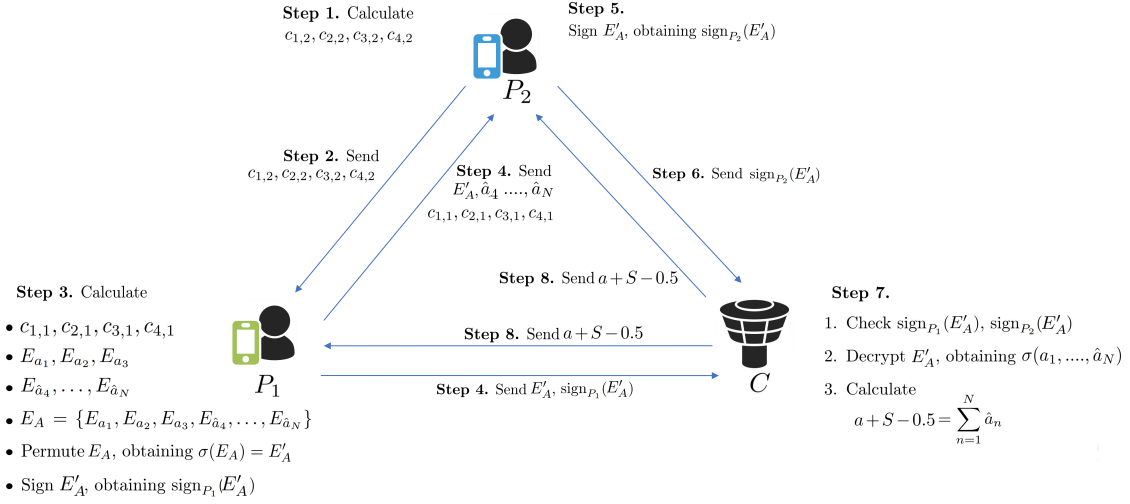}
\caption{Representation of the protocol steps}
    \label{fig:protocol}
\end{figure}

The proposed system is now described in higher detail. The steps described in this section are depicted in Figure \ref{fig:protocol}

We assume that a Control Center $C$ is willing to help calculate the distance between a user $P_1$ located at $(\lambda_1, \varphi_1)$ and another user $P_2$ located at $(\lambda_2, \varphi_2)$. We also assume that $C$ has an ElGamal private/public key pair $(s, y)$, where $y=g^s$, in a multiplicative group $\mathbb{Z}^*_q$. The public key is known by $P_1$ and $P_2$.

\subsection{Ciphertext generation}
During this phase, $P_1$ and $P_2$ are going to encrypt the values that $C$ will need in order to apply the Haversine formula and calculate distance. The process is shown in Table \ref{tab:cipher}, and as a result each of them obtains 5 ciphertexts necessary to calculate distance according to Equation \ref{eq:haver}. 


\begin{table}[htbp]
  \centering
  \caption{Ciphertext generation process}
    \def\arraystretch{1.5}
    \begin{tabular}{r|l|l|}
\cmidrule{2-3}         & \multicolumn{1}{c|}{$P_1$} & \multicolumn{1}{c|}{$P_2$} \\
         & \multicolumn{1}{c|}{Party 1} & \multicolumn{1}{c|}{Party 2} \\
\cmidrule{2-3}   
    1.   &     $c_{1,1} = (g^{r_{1,1}}, \cos \lambda_1 \cdot y^{r_{1,1}})$ &     $c_{1,2} = (g^{r_{1,2}}, \cos \lambda_2 \cdot y^{r_{1,2}})$ \\
    2.   &     $c_{2,1} = (g^{r_{2,1}}, \sin \lambda_1 \cdot y^{r_{2,1}})$ &     $c_{2,2} = (g^{r_{2,2}}, \sin \lambda_2 \cdot y^{r_{2,2}})$ \\
    3.   &     $c_{3,1} = (g^{r_{3,1}}, \cos \varphi_1 \cdot y^{r_{3,1}})$ &     $c_{3,2} = (g^{r_{3,2}}, \cos \varphi_2 \cdot y^{r_{3,2}})$ \\
    4.   &     $c_{4,1} = (g^{r_{4,1}}, \sin \varphi_1 \cdot y^{r_{4,1}})$ &     $c_{4,2} = (g^{r_{4,2}}, \sin \varphi_2 \cdot y^{r_{4,2}})$ \\
\cmidrule{2-3}    \end{tabular}%
  \label{tab:cipher}%
\end{table}%

\subsection{Ciphertext products} \label{products}
In this phase, the multiplicative homomorphic properties of ElGamal are going to be used to prepare data for the Haversine distance calculation. 

One of the agents must send his/her four ciphertexts to the other, who will perform the multiplications. Without loss of generality, we assume that $P_2$ sends ciphertexts and $P_1$ multiplies, but it could be the other way around with no consequences. So, at this point, $P_2$ sends $c_{1,2}$, $c_{2,2}$, $c_{3,2}$, $c_{4,2}$ and $P_1$ calculates:

\begin{align*} 
E_{11}= c_{1,1}\cdot c_{1,2} 
&= (g^{r_{1,1}}, \cos \lambda_1 \cdot y^{r_{1,1}}) \cdot (g^{r_{1,2}}, \cos \lambda_2 \cdot y^{r_{2,2}}) \\
&= (g^{r_{1,1}+r_{2,2}}, \cos \lambda_1 \cdot \cos \lambda_2 \cdot y^{r_{1,1}+r_{2,2}}) \\
\end{align*}

Due to the multiplicative homomorphic properties of ElGamal, multiplying these two ciphertexts, $P_1$ obtains an encrypted version of $\cos \lambda_1 \cdot \cos \lambda_2$. As shown in Equation \ref{eq:haver} in line The other products in the Haversine formula, are also computed by $P_1$:

\begin{align*} 
E_{22}= c_{2,1}\cdot c_{2,2} = (g^{r_{2,1}+r_{2,2}}, \sin \lambda_1 \cdot \sin \lambda_2 \cdot y^{r_{2,1}+r_{2,2}})\\
E_{33}= c_{3,1}\cdot c_{3,2} = (g^{r_{3,1}+r_{3,2}}, \cos \varphi_1 \cdot \cos \varphi_2 \cdot y^{r_{3,1}+r_{3,2}})\\
E_{44}= c_{4,1}\cdot c_{4,2} = (g^{r_{4,1}+r_{4,2}}, \sin \varphi_1 \cdot \sin \varphi_2 \cdot y^{r_{4,1}+r_{4,2}})
\end{align*}

Consequently, at the end of this phase, $P_1$ has four ciphertexts $(E_{11}, E_{22}, E_{33}, E_{44}$ with the encrypted version of all the products necessary to apply the Haversine formula. Let $E_{a_i}$ be the ciphertext corresponding to $a_i$, then:
\begin{enumerate}
 \item $E_{a_1}$ can be obtained by multiplying $E_{22}$ by $-0.5$ in the encrypted domain. 
    \item $E_{a_2}$ can be obtained by multiplying $E_{11}$ by $E_{44}$ and, then, multiplying the result again with the encrypted version of $-0.5$.
    \item $E_{a_3}$ can be obtained by multiplying $E_{11}$ by $E_{33}$ and, then, multiplying the result again with the encrypted version of $-0.5$.
\end{enumerate}


Now, the naïve approach to complete the protocol would be to transfer $E_{a_i}$ to $C$ such that $C$ could decrypt them, add them up yielding $a-0.5$, add up $0.5$ and, finally, obtaining $d$ using Equation \ref{eq:haver}. However, this approach would imply that $C$ has access to the decrypted form of the terms $a_i$. From those values, $C$ could try to obtain some information about the location of $P_1$ and $P_2$. Despite there would be more unknowns ($\lambda_1$, $\lambda_2$, $\varphi_1$ and $\varphi_2$) than equations (three), some information might be obtained from these values. Obviously, if $C$ is honest-but-curious, the naïve approach does not provide the necessary security and privacy requirements.


The following solution is proposed. First, we assume that $P_1$ is the only party sending data to $C$. If both $P_1$ and $P_2$ must send data to $C$, the protocol can be easily modified to allow it. For simplicity, we assume that both $P_1$ and $P_2$ are equally interested in knowing the distance and $P_1$ will not deviate from the protocol.

The only computation required to obtain the distance $d$ is the sum of all $a_i$. This sum will provide $a-0.5$ and, finally, $d$. $P_1$ can hide the relevant ciphertexts $E_{a_i}$ among a number of other encrypted values whose sum $S$ is known by $P_1$. Hence, $P_1$ chooses $N-3$ random numbers $\hat a_j$ for $j=4,\dots N$ and
\[
S= \sum_{j=4}^N \hat a_j,
\]
and encrypts them yielding $E_{\hat a_j}$ for $j=4,\dots N$. 

If unsigned integers are used in the encrypted domain, the signs must kept in a separate vector and send to $C$ to complete the computation. 

It must be taken into account, $a_1,a_2,a_3$ can differ in several orders of magnitude from each other. For this reason, the random numbers should be generated in the appropriate intervals in order to hide $a_i$ among similar values.


Now, $P_1$ has $N$ encrypted values $E_A=\{E_{a_1},E_{a_2},E_{a_3},E_{\hat a_4},\dots,E_{\hat a_N}\}$ that can be randomly permuted with a permutation function $\sigma$, obtaining $\sigma(E_A)=E_A^\prime$. Then, $E_A^\prime$ is sent to $C$. Upon receiving, $C$ can decrypt all of them as described in section \ref{sec:partial}, sum them all obtaining $a+S-0.5$. Then $P_1$ can subtract $S-0.5$ from $a+S-0.5$ obtaining $a$ and, finally, compute $d$.

The protocol can be easily adapted such that both $P_1$ and $P_2$ receive the same result. Both $P_1$ and $P_2$ can compute $a_1,a_2,a_3$ in the encrypted domain. $P_1$ can choose the random numbers $\hat a_j$ and their corresponding ciphertexts and send them (both as plaintext and ciphertext) to $P_2$, who can check that the encryption of all $\hat a_j$ is correct. Both $P_1$ and $P_2$ can sign the numbers $E_A$ and send them to $C$. Upon receiving of $E_A^\prime$ and, after checking the signatures of $E_A^\prime$ by $P_1$ and $P_2$, $C$ decrypts $E_A^\prime$, adds them up, and sends the result $a+S-0.5$ to both $P_1$ and $P_2$, who can complete the computation by subtracting $S-0.5$ (known to both) and finally obtain the distance $d$.

\section{Security analysis} \label{sec:security}
In this section, the security of the system is analyzed. First of all, the ElGamal cryptosystem is semantically secure under the Decisional Diffie-Hellman assumption. This means that a dishonest user cannot know if two different ciphertexts will result into the same cleartext after decryption.

We have three users in our scenario: $P_1$, $P_2$, and $C$. All ciphertexts are encrypted using  $C$'s public key. Consequently, only $C$ can decrypt any message.

 Considering the role of each user, three situations might threat the system security:
\begin{enumerate}
    \item \textbf{$P_1$ wants to learn the location of $P_2$}. The only contact $P_1$ has with the location data of $P_2$ is during the ciphertext products step (Section \ref{products}). At this point, $P_1$ receives the ciphertexts $c_{1,2}$, $c_{2,2}$, $c_{3,2}$, $c_{4,2}$. These ciphertexts contain trigonometric values of the latitude and longitude of $P_2$. However, since they are encrypted with the $C$' s secret key, $P_1$ cannot decrypt them and obtain any information about $P_2$'s location. 
    \item \textbf{$P_2$ wants to learn the location of $P_1$}. $P_2$ receives $E_A^\prime$ from $P_1$, the $N-3$ generated random values in cleartext ($\hat a_j$) and the ciphertexts $c_{1,1}$, $c_{2,1}$, $c_{3,1}$, $c_{4,1}$. First, $P_2$ encrypts the cleartexts of $\hat a_j$ and checks that they appear in $E_A^\prime$. This is done to ensure that $P_1$ has not maliciously altered the process. Consequently, $P_2$ knows that three remaining ciphertexts are $E_{a_1}, E_{a_2}, E_{a_3}$. However, since these values are encrypted with $C$'s public key, $P_2$ cannot learn any information about $P_1$'s latitude and longitude contained within. With the ciphertexts $c_{1,1}$, $c_{2,1}$, $c_{3,1}$, $c_{4,1}$, $P_2$ can compute $E_{a_1}, E_{a_2}, E_{a_3}$ also to check\footnote{Note that this step is not really necessary since $P_1$ could also cheat about $c_{1,1}$, $c_{2,1}$, $c_{3,1}$, $c_{4,1}$. If we omitted this step, some operations would be saved on $P_2$'s side. However, the protocol is fairer if both $P_1$ and $P_2$ obtain the same information from each other.} that all the values in $E_A^\prime$ are correct and compute a signature of $E_A^\prime$ to be sent to $C$.
    \item \textbf{$C$ wants to learn the location of $P_1$ or $P_2$}. $C$ can decrypt the permuted encrypted values $E_A^\prime$. However, these ciphertexts contain parts of the terms used in the Haversine formula mixed with some other rogue similar values. Consequently, $C$ cannot revert the process to learn the latitude and longitude of $P_1$ or $P_2$.


\end{enumerate}


\section{Performance analysis} \label{sec:performance}
In order to further analyze the protocol, we have studied the two elements that may have a higher impact on the performance: the communication and the computation costs. 

First of all, we have measured the communication cost based on the message complexity. In order words, we have counted how many messages a user must send during the protocol. Note that we consider sending a long message (for example a list of ciphertexts) as only one message. We do not deal with the fact that the underlying routing protocol might have to split them into several chunks. 
Consequently, there are five message transmissions:
\begin{itemize}


\item $P_2$ sends the ciphertexts $c_{1,2}$, $c_{2,2}$, $c_{3,2}$, $c_{4,2}$ to $P_1$.
\item $P_1$ sends the $N$ encrypted values  $E_A^\prime$, the ciphertexts $c_{1,1}$, $c_{2,1}$, $c_{3,1}$, $c_{4,1}$, and the random numbers $\hat a_j$ (as cleartext) to $P_2$.
\item $P_1$ sends the $N$ encrypted values  $E_A^\prime$  with his/her signature to $C$.
\item $P_2$ checks the values received form $P_1$ and, if they are correct, sends his/her signature of $E_A^\prime$ to $C$.
\item $C$ broadcasts the resulting calculation $a+S-0.5$ to $P_1$ and $P_2$.

\end{itemize}


Regarding the computation cost, in our protocol this is inherent to cryptographic operations, specially modular exponentiations. Analyzing the protocol, the total number of exponentiations is:

\begin{itemize}
    \item Encryption under ElGamal requires two exponentiations, one for the first part of the ciphertext $g^r$, and the other for the second part $m\cdot y^r$. Although these exponentiations are independent of the message and can be computed ahead of time if need be, for our scenario we will treat them as if calculated during execution time. Consequently, since $P_1$ and $P_2$ have to generate the four ciphertexts of Table \ref{tab:cipher}, each of them computes 8 exponentiations.
    \item $P_1$ has to encrypt $N - 3$ values to hide the relevant ciphertexts. This results into $2(N-3)$ exponentiations. $P_2$ also encrypts the same values to check that $E_A^\prime$ received from $P_1$ are correct.
    \item Regarding the signatures, each ElGamal signature requires one exponentiation. $P_1$ and $P_2$ have to each sign one message, hence they compute one more exponentiation. $C$ has to verify both signatures, resulting into four exponentiations.
    \item Decryption under ElGamal only requires one exponentiation. Since $C$ has to decrypt $N$ ciphertexts, this requires $N$ more exponentiations.
\end{itemize}

We can assume that the number of exponentiations performed by $C$ is irrelevant, since the Control Centre must be designed with enough computation power. $P_1$ and $P_2$ may however be performing their operations on a more restricted device,  like an smartphone with GPS location. 

The work presented in \cite{canard2012} analyzes the cost of executing a cryptographic protocol on a restricted device such a smartphone. For example, for a Samsung Galaxy S2 smartphone with a Dual-core Exynos 4210 1.2 GHz processor ARM Cortex-A9 with the Android OS, v2.3 (Gingerbread), the authors state that an exponentiation takes 42 ms. In order to compare it to a more modern device, we have performed the same test for a OnePlus 5 smartphone with a Qualcomm® Snapdragon™ 835 Octa-core, 10nm, up to 2.45 GHz processor with the OxygenOS based on Android™ Nougat. In this device, an exponentiation takes 0.19 ms. 

$P_1$ and $P_2$ have to each compute $8+2(N-3)+1 = 2N-3$ exponentiations. Consequently, the computation time depends on the parameter $N$. For higher $N$ values, users increase $C$'s uncertainty but reduce performance. For example, assuming $N=100$, each user's computations take 8.274 s approximately in the Samsung Galaxy S2 smartphone, and 37.43 ms in the OnePlus 5 smartphone.

\section{Conclusions} \label{sec:conclusions}
With the extended use of mobile devices and applications, there is a growing number of mobile services which requiere location data. Although users can obtain many benefits of these services, continuously revealing the exact location to an external entity is a threat for the involved parties. 

For this reason, in this paper we propose a secure multi-party protocol which allows to calculate distances between users, but without requesting location disclosure. In our protocol, using ElGamal cryptosystem and its homomorphic properties, it is not necessary to reveal the latitude and longitude in order to obtain the distance. 

We have analyzed our protocol in terms of security and performance. The security analysis shows that none of the involved entities can learn any information about the location of the users. The performance analysis shows that the computation and communication costs are completely reasonable for an equipment with similar capabilities to a smartphone device. 

\section*{Acknowledgments}
This work is partly funded by the Spanish Government through grant RTI2018-095094-B-C22 ``CONSENT''.

\bibliographystyle{alpha}
\newcommand{\etalchar}[1]{$^{#1}$}

\end{document}